\documentclass[preprint2]{aastex}

\shortauthors{R. Kehoe et al.}
\shorttitle{Untriggered Optical Burst Search}

\received{2002 February 20}
\begin{document}

\title{An Untriggered Search for Optical Bursts }

\author{Robert Kehoe\altaffilmark{1,4}, Carl Akerlof\altaffilmark{1}, Richard Balsano\altaffilmark{2}, 
Jeff Bloch\altaffilmark{2}, Don Casperson\altaffilmark{2}, Sandra Fletcher\altaffilmark{2}, 
Galen Gisler\altaffilmark{2}, 
Brian Lee\altaffilmark{1,5}, Stuart Marshall\altaffilmark{3}, Timothy McKay\altaffilmark{1}, 
Eli Rykoff\altaffilmark{1}, Donald Smith\altaffilmark{1},
Tom Vestrand\altaffilmark{2} and Jim Wren\altaffilmark{2}}

\altaffiltext{1}{University of Michigan, Ann Arbor, MI 48109}
\altaffiltext{2}{Los Alamos National Laboratory, Los Alamos, NM 87545}
\altaffiltext{3}{Lawrence Livermore National Laboratory, Livermore, CA 94550}
\altaffiltext{4}{Michigan State University, East Lansing, MI 48824}
\altaffiltext{5}{Fermi National Accelerator Laboratory, Batavia, IL 60510}

\begin{abstract}
We present an untriggered search for optical bursts with the ROTSE-I telephoto
array. Observations were taken which monitor an effective 256 square degree
field continuously over 125 hours to \( m_{ROTSE}=15.7 \). The uniquely large
field, moderate limiting magnitude and fast cadence of $\sim$10 minutes permits
transient searches in a new region of sensitivity.  
Our search reveals no candidate events. To 
quantify this result, we simulate potential optical bursts with peak
magnitude, \( m_{p} \), at t=10 s, which fade as 
\( f=\left(\frac{t}{t_{0}}\right) ^{\alpha _{t}} \), where \( \alpha_t < 0 \).
Simple estimates based on observational evidence indicate that a search of this
sensitivity begins to probe the possible region occupied by GRB orphan afterglows.
Our observing protocol and image sensitivity result in a broad region of high
detection efficiency for light curves to the bright and slowly varying side
of a boundary running from \( [\alpha _{t},m_{p}]=[-2.0,6.0] \) to \( [-0.3,13.2] \).
Within this region, the integrated rate of brief optical bursts is less than
\( 1.1\times 10^{-8}~{\rm s}^{-1}~{\rm deg}^{-2} \). At $\sim$22 times the observed
GRB rate from BATSE, this suggests a limit on 
\( \frac{\theta _{opt}}{\theta _{\gamma }}\lesssim 5 \) where \( \theta_{opt} \) 
and \( \theta_{\gamma} \) are the optical and gamma-ray collimation angles, 
respectively.
Several effects might explain the absence of optical bursts, and a search of the 
kind described here but more sensitive by about 4 magnitudes should offer a more 
definitive probe.
\end{abstract}



\section*{Motivation}

Gamma-ray bursts remain one of the great mysteries in astrophysics. 
Although there have been some concrete measurements
of the energetics of some bursts through redshift determination (eg. \cite{metzger97},
\cite{kulkarni98}), there is little firm knowledge of how the energy is produced. In
fact, the total production is still uncertain by approximately 2 or 3 orders
of magnitude because of the unknown level of postulated collimated jets. 
Additionally, despite recent advances in multi-wavelength detection strategies, the 
total number of GRBs studied optically remains small.  While various observations
have placed the internal-external shock scenario on a relatively firm footing
(eg. \cite{akerlof99}, \cite{sari99}) for some bursts, it is not verified for most. 
Bright optical bursts are the expected signature of reverse external shocks, but 
have been ruled out for several gamma-ray bursts 
\cite{akerlof00}, \cite{kehoe01}, \cite{park99}, which argues against a uniform 
behavior. 

Gamma-ray bursts are believed to emit synchrotron or inverse
Compton radiation from material moving at ultra-relativistic velocities. The
resultant strong Lorentz beaming of the emission will decrease as the shocked
material slows down. If the ultra-relativistic bulk flow is a collimated jet, 
radiation at wavelengths
longer than gamma rays, believed to be produced by this slower material, will
be emitted through a larger solid angle \cite{rhoads97}. This suggests a population of 
orphan optical bursts with timescales similar to GRBs, but more
frequent and with no gamma-ray signature.

With the unfortunate demise of the Compton Gamma-Ray Observatory, the rate of GRB 
detections overall has 
decreased dramatically.  The physics of jets may prove valuable in the 
ongoing effort to observe multiwavelength emission from their progenitors and in a
wider variety of conditions than studied so far.  Conversely, the study of the 
relative rates of gamma-ray and ``orphan'' optical bursts
may uniquely probe the open question of collimation. Whether useful
constraints can be obtained in this way is currently
the subject of some debate, some holding the view that strong limits are 
possible from such a measurement (eg. \cite{rhoads97}). 
On the other hand, \cite{dalal01}
have asserted these efforts are unable to measure the collimation angle.  However,
in \cite{dalal01} only very late, slow cadence,
and narrow-field searches were discussed. In addition, the estimates of optical
lightcurves are based heavily on an interpretation of a model which has yet
to be confirmed for GRB's. 

Observational evidence for bursts at longer wavelengths than gamma-rays 
is at present very sketchy. One search for the expected higher rate of 
X-ray afterglows has been performed \cite{grindlay00} with negative results.
On the other hand, mysterious optical transients have been observed in deep,
narrow-field searches (eg. \cite{tyson}). Although selected based on
spectral criteria rather than photometric (lightcurve) critera as in our
paper, the SNe optical search of \cite{sdss} has recently revealed a new type of 
AGN at 17th magnitude exhibiting characteristics reminiscent of GRB optical afterglows.
This is close to the ROTSE-I detection threshold, and as a background \cite{galyam}
will have to be understood for any GRB orphans search.

A substantial complication to these kinds of searches arises from the fact that the 
rates of rapidly varying optical transients are poorly understood.
Study of eruptive variables with timescales of variation of order $\sim$1h has been 
limited to searches with very non-uniform sky coverage. The catalog of known rapidly 
eclipsing and pulsating systems has a similar limitation. The rates of
burst-like AGN activity, such as from SDSS J124602.54+011318.8, are
also very poorly understood.  Aside from being interesting in their own right, these
transients are sources of background for GRB optical counterpart searches, 
whether triggered or untriggered.

\section*{Method}

For all of these reasons, we have devised an optical burst search strategy based
almost solely on considerations taken directly from data. Enough GRB late optical
afterglows have been measured to indicate that lightcurves of the form
\( f=\left( \frac{t}{t_{0}}\right) ^{\alpha _{t}} \), with \( \alpha _{t} \sim -1 \),
are typical. When extrapolated back to times nearer the burst phase,
the observed afterglow lightcurve of GRB 970228, which is fairly typical of
those observed so far, indicates emission brighter than 16th magnitude
for the first 30 minutes. While the non-detections of optical bursts imply this
is not the typical scenario \cite{kehoe01}, \cite{akerlof00}, \cite{park99}, the observed
optical burst from GRB 990123 was actually brighter than this extrapolation
would indicate. This sensitivity is attainable in 3 minute exposures with ROTSE-I
\cite{kehoe00}, and it places the observation of an optical burst lightcurve brighter
than this limit over 30 minutes just within the telescope's capability.  We therefore
take this lightcurve signature as our fundamental search criterion.

Because this search is limited by field area covered and integrated observation time,
it makes sense to maximize both in our study. The hypothesized 30 minute window 
of detectibility for ROTSE-I sets the number of different
pointings we can accomplish while we expect an optical burst to be visible
which, in turn, sets the cadence of our observations. We will discuss
this more in the Observations section, but ROTSE-I can accomplish 2 pointings in 
the allotted time. In general, for a rapidly fading lightcurve,
the cadence of a wide-field search must increase with decreasing aperture.
This permits fewer pointings by the smaller telescope but this is compensated for
by the larger field of view.  
Despite the modest sensitivity of ROTSE-I, we are aided by its relatively large etendue,
\( D^{2}\Omega  \), a measure of sensitivity for such searches. This product
varies relatively little over a wide range of instruments because the useful
solid angle for fixed focal plane area decreases as \( D^{-2} \) as
the aperture, $D$, increases.  Thus, 
under the assumptions made in this paper, ROTSE-I is roughly equivalent to a 0.5m, 
\textit{f/1.8} system.  At the
BATSE fluence sensitivity of \( 10^{-8} \) to \( 10^{-7}~{\rm ergs}/~{\rm cm}^{2} \), 
there are approximately two bursts per \( 4\pi  \) steradians per day. If the optical
emission fills a cone 5 times wider than that for the gamma-rays, then a detectable
optical burst potentially occurs once per 160 hours in a single ROTSE-I field
of 256 square degrees. 

The uniquely large field and fast cadence
of the observations taken for this analysis push transient searches in a very
different part of observational phase space than have been pursued before. To both
tune our search strategy and estimate our sensitivity to the sought after signature, we
have employed a simple Monte Carlo simulation.  This simulation has three main functions:
(1) reproduce the photometric behavior of ROTSE-I measurements, (2) explore our search 
efficiency for a range of input optical burst lightcurve parameters, and (3) quantify the 
search sensitivity given the observing protocols and limiting magnitudes of the actual 
data taken.  We ignore in this simulation model dependent assumptions about relativistic
beaming or spectral evolution so that our result can be interpreted as generally as possible.
We will describe this simulation more in the Analysis section.

\section*{Observations}

Our search for optical bursts utilizes data taken with the ROTSE-I telephoto
array \cite{kehoe00}, which consists of four telescopes arranged in a 
\( 2\times 2 \) configuration, each with an 8 degree field of view and 14{}''
pixels.  The major challenge in our analysis is the successful rejection of 
$\sim$\( 1.5\times 10^{5} \)
non-variable sources for over 2000 observations in each pointing, as well as
suppression of astrophysical transients and instrumental backgrounds.
At a bare minimum, any candidate for an optical burst must be well-detected in
at least two observations.  This stipulation, in conjunction with the 30 minute window 
constraint, largely dictates our observational approach.

We have taken the data used in this analysis in two main pointing modes, 
\textit{stare} and \textit{switch}. The first mode consists of repeated images 
of a single field each night with no repointings
to other locations. This \textit{stare} data was acquired in two different periods.
In mid-December of 1999, two separate fields were monitored on different days,
while in mid-April, we covered another field extensively, originally to monitor
the X-ray nova XTE J1118+480 \cite{wren01}. Camera `d' was not operational during
half of these observing nights, and camera `c' was inoperable on one. The total
\textit{stare} data set comprises an effective monitoring of 256 square degrees
in 80 s exposures for 47.5 hours. The observation dates and coordinates for
these fields, and the total area covered, are given in Table 1.

While the \textit{stare} data provides excellent temporal coverage of any transients
which may occur, the field-of-view probed during our crucial 30 minute window
is unduly restricted. As mentioned above, the 
finite visibility window of the hypothesized optical bursts dictates 
that we maximize the area covered
in the allotted time. With ROTSE-I, we can fit two pointings into this timeframe.
We employ a protocol in which five 80 s exposures are taken at each pointing, followed
by a repointing at a second location where another five exposures are taken.
The telescope then returns to the first field and restarts. This leaves blocks
of five 80 s exposures with 7 minute gaps for each of the two pointings. One
peculiarity of this \textit{switch} mode occurs when we co-add pairs of images later
-- namely, each cycle has an un-coadded fifth image. This extra image alternates
between preceding or following the two co-adds. We have chosen two of our standard
sky patrol fields which pass through the zenith during the late summer and which
have a galactic latitude greater then 20 degrees to avoid Galactic extinction
and overcrowding. The data were taken from late August through early October 2000
during periods chosen to avoid bright moonlight (see Table 1).

For the observations presented in this analysis we instituted a dithering observing
procedure to suppress backgrounds stationary on the CCD.
This involved a Z-shaped rastering by about 2.5{}' on a side ($\sim$10
pixels) during the observing sequence to relocate bad pixels in celestial
coordinates in consecutive frames.

\section*{Data Reduction}

Reduction of the raw images taken for this study involves an initial calibration
of the images, through a co-addition and recalibration phase, to lightcurve 
construction.
This analysis generated 14,000 raw images which were processed to approximately
8000 epochs and \( 2.5\times 10^{8} \) photometric measurements. The backgrounds
and processing time incurred necessitated data reduction as the images were taken,
while adding the ability to diagnose and, if possible, correct for various issues
of image quality.

The initial calibration of raw images is described in \cite{kehoe00} and involves
three main steps: correction of images, source extraction, and astrometric and
photometric calibration. 
Images are corrected through the subtraction of dark current and flat-fielding.  The 
darks, flats and bad pixel lists are generated at the beginning of the three major 
periods of data collection.  The corrected images are analyzed using SExtractor 
\cite{sextr}
to produce object catalogs. To speed up processing, we have optimized the choice
of clustering parameters while not significantly degrading the ability to extract
sources in crowded environments. When an object aperture contains a bad pixel,
we flag the object in that observation as bad. The photometric and astrometric 
calibration involves matching
the resultant list of sources to the Hipparchos catalog \cite{tycho}. 
As will be described in the Analysis section, our search depends on the statistical 
study of lightcurve data.  For our final calculated search efficiencies to be 
correct, we must reproduce the observed photometric fluctuations seen in actual ROTSE-I 
data.  We find that
an irreducible 0.01 mag fluctuation, in addition to the normal statistical one,
reproduces the observed behavior well.  We assign this as the minimum systematic
error for all observations.

\subsection*{Final Image Processing}

Once the astrometry for each frame is established, we can co-add images in pairs to 
improve our sensitivity.  
We do this instead of taking exposures of twice the length to increase the dynamic range
available in our 14-bit CCDs, to help suppress bad pixel and other instrumental 
backgrounds, and to permit a later augmentation of the search at twice the temporal 
granularity.  For each field per night (ie. 4 for \textit{stare}, 8 for 
\textit{switch}), we independently number frames starting at `1' for the first image.  
We co-add each even numbered frame to its immediately preceding odd numbered frame if 
it is adjacent in time.
The even numbered frame is mapped to the odd numbered frame using a bilinear
interpolation of pixel intensities to best avoid position resolution degradation
and image artifact creation.  During the co-addition, intensities
of known bad pixels in the first and second images are replaced with the median
of their 8 neighbors and are given zero weight. 
Pixels in the final co-added image with weight $<$ 2.0 are rescaled to homogenize the
image.  Since the first and second images
are dithered with respect to each other, bad pixel locations generally
have correct image data from one of the two images. As a further precaution, once
the final co-added image is itself clustered and calibrated as described above,
found objects are flagged if a bad pixel falls within their 5 pixel wide aperture. 
Typical
image sensitivities of \( m_{ROTSE}=15.7 \) were obtained for co-added images. 

The next step of the processing chain involves removing images which
exhibit instrumental problems. Images must satisfy a \( 5\sigma  \)
limiting magnitude of at least 14.75, regardless of whether they are co-adds.
This generally removes epochs from early or late in a night which are affected
by twilight, although it also rejects images taken when weather is not of
good observation quality. The position resolution of an image must also
be less than 0.15 pixel ($\sim$2 arcsec). This removes images
which did not co-add well, due usually to some large obstruction (eg. a tree late 
in the evening as the field moves towards the horizon) 
or rare slipping of the mount clutch which compromises the astrometric warp map.

The last step before lightcurve construction involves the elimination from the
data sample of epochs or whole nights exhibiting instrumental problems. For
nights with \( N \) operational cameras, we omit epochs when fewer than $N-1$ cameras
are taking observation quality images. We require whole nights to have at least
two contiguous hours of images that pass quality cuts, where contiguous means that
no two consecutive epochs are missing. The surviving nights generally exhibit
very stable quality over those observations we included. To quantify our observations,
we calculate for each good night the median limiting magnitude of the images
used, and the total observing time. These are indicated in Figure 1.

\subsection*{Lightcurve Construction}

For each camera, the good object lists for each night are matched to provide
lightcurves for all detected sources. We remove any object which is not seen
in at least one consecutive pair of observations, while we retain all isolated
observations of objects passing this selection. To avoid suspect photometry,
intermittent sources arising from deblending fluctuations, or moving objects,
we flag a source's observations when the measured position is more than one
half pixel from the mean source location.

After we have created a filtered list of calibrated lightcurves, we must perform
an internal correction to remove systematic photometric mismeasurements due
to such difficulties as the presence of thin haze over a portion of a frame,
or increased vignetting due to a sticking shutter. More serious problems
occur when power-lines or trees cross an image or when shutters obstruct a portion
of an image. In each of these cases, we determine either a small relative photometric
correction and systematic error, or in more extreme situations flags to indicate
the problem and exclude the image area. We begin by selecting a set of good
template sources in each matched list. For each source to enter this list, we
count those observations which have good photometry. If these criteria are satisfied
for more than 75\% of the epochs in which the source location was imaged, the
object becomes a template source and we calculate its median magnitude in good
observations. Each image is then divided into 100 pixel square \textit{subtiles},
and photometric offsets for each template source having statistical error less 
than 0.1 mag in these subtiles are calculated. The number of template sources, 
\( n_{i} \),
within a typical subtile, \( i \), is around 50. A relative photometry map
is then calculated to be the median of these offsets, \( o_{i} \), for each
subtile. The standard deviation of the offsets, \( \sigma _{o,i} \), which
for good subtiles is typically around 0.03 magnitudes, is used to calculate
a systematic error \( \varepsilon _{i} \) on \( o_{i} \). In addition, subtiles
are classified as bad if \( n_{i}<5 \), or if \( \sigma _{o,i}>0.1 \) magnitude.
These selections remove obstructions and large photometric gradients due to
out-of-focus foreground objects, which constituted approximately 2\% of the
data sample. Each observation of each object is then corrected by the \( o_{i} \)
using a bilinear interpolation, and the \( \varepsilon _{i} \) is added in
quadrature to that observation's systematic error. If the subtile is flagged
as bad, then the object's observations are likewise flagged.

The lightcurve analysis proceeds from those observations of a particular source
which pass basic quality cuts. The source must neither be saturated, nor near
an image boundary. The aperture magnitude must be properly measured. The position
of the source in good observations must be less than 0.5 pixels from the source's
mean coordinates. The inefficiency incurred for these selections is negligible.
There must also be no known bad pixels within the source aperture, and the subtile
containing the source must have a well-measured photometric offset. The former 
inefficiency is approximately 10\%, while the latter is 2\% as stated earlier.

\section*{Analysis}

\subsection*{Lightcurve Selection}

Based on our observationally motivated model, we look for lightcurves which may
have timescales for the brightest optical emission which is similar to the
gamma-ray emission of a GRB.  In addition, we may expect a decaying lightcurve
with $\alpha_{t}<0$.  This signature is unlike any other known type of optical
transient, and our lightcurve selection exploits this. We begin by cutting on simple
statistical variables designed to efficiently identify a fast-rising transient
with a power-law decay, while removing backgrounds. These backgrounds come in
two kinds. The first originate from astrophysical processes such as low amplitude
pulsating and eclipsing systems, irregular fast variables, flare stars, and
asteroids. The second kind is dominated by various instrumental transients like those
resulting from unregistered bad pixels, poor photometry, or moderate mount clutch
failures not identified by our image quality cuts.

In order to obtain an \textit{a priori} selection, we first studied the instrumental
background by performing a grid search on the April 9th \textit{stare} data
of camera `b' ($2\%$ of the total search sample) in terms of the maximum variation 
amplitude, \( \Delta  \), and the significance of that variation, 
\( \sigma _{\Delta } \). We took as comparison the 13 eclipsing and pulsating systems 
we observe in this sample down to variation amplitudes
of approximately 0.1 magnitude. We explored the region of \( \Delta =[0.1..0.25] \)
and \( \sigma _{\Delta }=[2.0..5.0] \), and we selected cuts of \( \Delta >0.1 \)
and \( \sigma _{\Delta }>5.0 \) which minimized the total background while
retaining substantial efficiency for real, low amplitude variables.  

These variables generally come in two main types: pulsating stars such as the
five indicated in Figure 2, and eclipsing systems such as the three illustrated in 
Figure 3.  Of these 8 variables, only one is previously identified as such:
the Algol binary BS UMa (ie. ROTSE1 J112541.62+423448.8) \cite{bsuma}.
The rest are low amplitude variables difficult 
to identify in photographic plates.  ROTSE1 J112037.63+392100.4 and 
ROTSE1 J113536.74+384557.4 may be a $\delta$ Scu and an RRc, respectively, but we
have not fully classified the variables in this sample at this time.
However, averaging over four fields, we estimate the rate for variables 
with amplitude $> 0.1$ mag and period $< 1$d to be of order
$0.1{\rm deg}^{-2}$ for eclipsing systems and $0.2{\rm deg}^{-2}$ for pulsating systems.  
Several variables with longer
timescale variation (ie. $~0.7-3$ d) are also observed, and the approximate 
rate of these is $0.2{\rm deg}^{-2}$.  Note that the data is out of the galactic plane.

We further investigated the instrumental background behavior in \textit{switch} data 
taken in July, 2000. These fields are not included in the search due to a substantial
worsening of the problem with the tracking of the mount.  This was later almost 
completely fixed, greatly reducing the background in subsequent \textit{switch} data. 
The remaining backgrounds largely consisted of very stable sources with one 
anomalous photometric measurement. Cutting on a \( \chi ^{2} \) calculated 
from observations not including the most significant variation, \( \chi _{c}^{2} \), 
was effective against these.  A candidate flare-star, ROTSE1 J160542.74+350016.3, was 
observed in outburst twice in this July data as shown in Figure 4.

We examined the effect of this selection on our hypothesized optical burst signal.
We simulated the expected GRB signature with a simple Monte Carlo generating a variety 
of lightcurves with no emission during the first 10 s, and then peaking at \( m_{p} \) 
from 6.0 to 15.0. A power-law fading with \( \alpha _{t} \) from -0.05 through -3.0 
followed thereafter. A cut of \( \chi _{c}^{2}>3.0 \) is very efficient for the 
simulated bursts and removes most backgrounds passing our other cuts. Note that the 
calculation of \( \chi _{c}^{2} \) requires an object be seen in at least three good 
observations. 
This in turn means our search will be insensitive to variations lasting less than about 
7 minutes.
All of the simulated optical bursts passing our \( \chi _{c}^{2} \) cut 
also had a \( \Delta >0.5 \) magnitude.  We tighten our selection accordingly to 
remove most of the remaining low amplitude periodic variables seen in the data 
background samples.  Thus our lightcurve cuts require an overall significant variation 
with \( \Delta >0.5 \) and \( \sigma _{\Delta }>5.0 \), and substantial variability
in the rest of the lightcurve with \( \chi _{c}^{2}>3.0 \).  Although designed with a 
simple power-law time decay in mind, these cuts are efficient for a wide range of 
optical burst lightcurves.

At this stage, our candidate sample was dominated by three major sources of
background. The first consisted largely of bright, moderate amplitude pulsating
and eclipsing systems, most of which have not yet been cataloged. The second
category consisted of non-varying sources present in most observations but poorly
measured in more than one observation. For the most part, these latter are not
flare stars. These categories of candidate are easily measured on other nights
and so rejected. Given the results of our simulated lightcurves, these are highly
unlikely to correspond to GRB optical counterparts.
The last category arises from the fact that the outer region of each camera
field has relative photometry corrections which start to break down where interpolation
becomes extrapolation. We discard these regions in our search, and we thereby incur 
a 2.5\% inefficiency.

Figure 5 indicates, for a co-add sensitivity of 15.7 magnitude, the 50\% efficiency
contour for our simulated bursts after our final lightcurve cuts.  GRB 970228 and 
GRB 990123 are shown for comparison. We see that
the selection is efficient for simulated bursts in a particular subset of the
\( m_{p} \) vs. \( \alpha _{t} \) parameter space. The boundary of this region
is relatively sharp, and moves vertically as a night's typical
image sensitivities increase. The increased sensitivity to steep lightcurves
with the \textit{stare} protocol results from the uninterrupted nature of the
observing.

\subsection*{Search Results}

After the selection mentioned in the previous section, 50 candidate lightcurves 
remain.  Twenty six of these candidates are located along a CCD column in camera 
`c' with improper charge transfer and so rejected.  Hand-scanning and comparison to 
images without candidates revealed that 12 of the remaining 24 lightcurves were 
due to bad pixels. On three nights (000824, 000903,
000906), there was an intermittent 15\% increase in the number of hot pixels
observed in data. The reason for this behavior is not understood. Although it
affects a small portion of image real-estate, these pixels were not properly
dark-subtracted, flagged or removed in co-addition. One set of candidates happens
when two bad pixels are near enough that they land on the
same \( \alpha  \), \( \delta  \) in consecutive frames. The other candidates
consist of a hot pixel landing between two nearby bright stars, overlapping a 
star or stars
just below reliable detection, or following a satellite trail in the next image. 

Ten of the remaining 12 candidates are within 15{}'' ($\sim$1 pixel) of objects in the 
USNO A2.0 catalog \cite{usno} brighter than 17th mag. in $R$. By rejecting these, we 
incur an inefficiency dominated 
by the loss of the image region contained in $\sim 6$ pixel area around all of the
USNO source positions. Since the largest occupancy per camera is about 50,000 stars,  
this cut incurs a 7\% inefficiency.

The last two candidates clearly move slowly in our images. These match
the expected coordinates of the asteroids (1719) Jens and (17274) 2000 LC16,
and we reject them.  Thus, we find no optical bursts in this study.

Our data sample was taken at several different sensitivities and both \textit{stare}
and \textit{switch} pointing modes, and this will alter the sensitivity of
our search from that shown in Figure 5. Using the simple Monte Carlo already
described, we generated over 20 million bursts folding in the fractions with
co-add sensitivities and pointing mode given in Figure 1. A region of 
\( [\alpha _{t},m_{p}] \)
parameter space is efficiently accepted which lies on the bright and long side
of a rough line between {[}-2.0,6.0{]} and {[}-0.3,13.2{]}. This result is shown
in Figure 6. A final additional inefficiency of 20\% for our selection comes
from three main sources: bad pixel removal (\( \varepsilon =0.9 \)), bad image
regions (\( \varepsilon =0.98 \)), removal of previously known sources 
(\( \varepsilon =0.93 \)),
and removal of field edges (\( \varepsilon =0.975 \)).

\section*{Discussion}

We have performed the most extensive untriggered search for optical bursts in
wide-field data to date. After study of \( 1331.4~{\rm deg}^{2}~{\rm days} \), 
we find no candidates down to a typical limiting sensitivity of \( m_{ROTSE}=15.7 \).
This sensitivity is of the order necessary to study the potential rates for optical 
bursts from GRBs based on simple beaming hypotheses. At a maximum efficiency of 
80\%, we accept transients which are brighter and longer than a boundary running from 
\( [\alpha _{t},m_{p}]=[-2.0,6.0] \) to \( [-0.3,13.2] \). In this region, we 
therefore rule out an integrated optical burst rate greater than 
\( 1.1\times 10^{-8}~{\rm s}^{-1}~{\rm deg}^{-2} \). At a rate ten times greater, 
we reject bursts bounded by \( [\alpha _{t},m_{p}]=[-2.6,6.0] \) to \( [-0.3,14.3] \).

As there is no empirical information about the optical properties of orphan afterglows,
any interpretation of this result in the context of GRB beaming is uncertain.
If we consider that the optical emission is collimated into some jet angle, 
$\theta_{opt}$, which may be different than the gamma-ray collimation angle, 
$\theta_{\gamma}$, then the relative rate of gamma-ray to optical bursts may indicate
something about the relative magnitudes of these angles.  Compared to the observed 
GRB rate from BATSE of \( 5\times 10^{-10}~{\rm s}^{-1}~{\rm deg}^{-2} \),
our limit on optical burst rates translates into a limit on 
\( \frac{\theta _{opt}}{\theta _{\gamma }}\lesssim 5 \).

This result seems to conflict with the much higher rates predicted in the analysis
of \cite{frail01}. There may be several reasons for this. We have assumed that all 
of the
bursts observed by BATSE would have produced optical bursts of sufficient intensity
to be observed above 16th magnitude. The relationship of optical emission during
this early phase to gamma-ray emission is not known well enough to rigorously
justify this assumption. In this case, however, a search of the kind described in
this analysis but deeper
by 2-4 magnitudes should provide a more incisive probe, given the distribution
of GRB fluences relative to those we have discussed in this paper. 

Although most GRBs with 
both X-ray and radio afterglows seem to have optical counterparts, it is still 
possible that obscuration at the source is dimming a large number
of the optical bursts we would otherwise see. For instance, there is evidence
for as much as 5 mag extinction for GRB 970828 \cite{groot98}.  Ultimately, the
question of extinction at the source is still tied up with the identity of the
progenitor.  In collapsar models \cite{woosley93}, substantial extinctions could result.

Lastly, we have ignored the specifics of how the changing Lorentz factor of the burst 
ejecta affects the very early optical emission. 
This behavior is complex, and very model dependent, but for very off-axis viewing
the observed optical flux is likely to peak later, and at a lower value (eg. 
\cite{granot02}), than 
might be suggested by extrapolating back from the late afterglow.  
Observationally, many optical afterglows have exhibited breaks in their lightcurves
which may indicate analogous behavior for nearly on-axis viewing \cite{frail01}.  
In a couple of cases, the early lightcurve exhibits a time decay index of $\sim -0.8$
(\cite{stanek99}, \cite{stanek01}) which puts them in a relatively sensitive region of 
Figure 6.  Any optical burst with such a slope and brighter than about 12th mag at 
it's peak would have been observed, and bursts with peak brightnesses of $\sim 13$th 
mag would be detected with lower efficiency.
As one moves more off-axis, the slope at early times becomes shallower until 
eventually it may rise to the break point.  The suppression of this early emission will
result in a decreased ability to observe these bursts.  However, it should be kept in 
mind that, regardless of when the peak occurs relative to the initial event, our 
search will be 
sensitive to these lightcurves if the emission is brighter than 16th mag in a way
similar to the optical bursts we have simulated (ie. detection in at least three images with overal \( \Delta >0.5 \)).
In any case, a limiting magnitude deeper by $\sim$ 4 mag would substantially improve 
the chances for detection, especially in light of observed GRB optical afterglow 
brightnesses.

This search technique can be improved and extended in several ways. The main
source of background was instrumental. To consider much larger numbers of observations, 
it is important to reduce the number of bad pixels and shutter malfunctions, and 
avoid visual obstructions.  Another improvement would be to increase
the sophistication of the relative photometry calibration. Most
importantly, the next step for an untriggered optical burst search
would be to consider deeper fields, such as available with the
ROTSE-III 0.5m telescopes.  This kind of search will likely be
looking for more gradual optical variation than sought in this
paper. For robust identification, it may be necessary to obtain
spectral verification of candidate orphans which are more likely to be
found at the sensitivity of R0TSE-III. The search must be made
immune to the kind of background engendered in SDSS
J124602.54+011318.8, in particular.

This analysis has implications for our ability to extend our GRB studies. We 
have already used the techniques described here to
search for optical counterparts for several archival BATSE triggered bursts
\cite{kehoe01}. In addition, they are a crucial step towards our goal of
near real-time detection with the ROTSE-III telescopes the optical transients 
associated with GRB triggers from satellite-based gamma-ray observations.

\begin{acknowledgements}
ROTSE is supported by NASA under SR\&T grant NAG5-5101, the NSF under grants
AST 97-03282 and AST 99-70818, the Research Corporation, the University of Michigan,
and the Planetary Society. Work performed by LANL is supported by the Department
of Energy under contract W-7405-ENG-36. Work performed at LLNL is supported
by the Department of Energy under contract W-7405-ENG-48.
\end{acknowledgements}

\clearpage

\begin{figure}
\vspace{0.3cm}
{\par\centering \resizebox*{10cm}{!}{\includegraphics{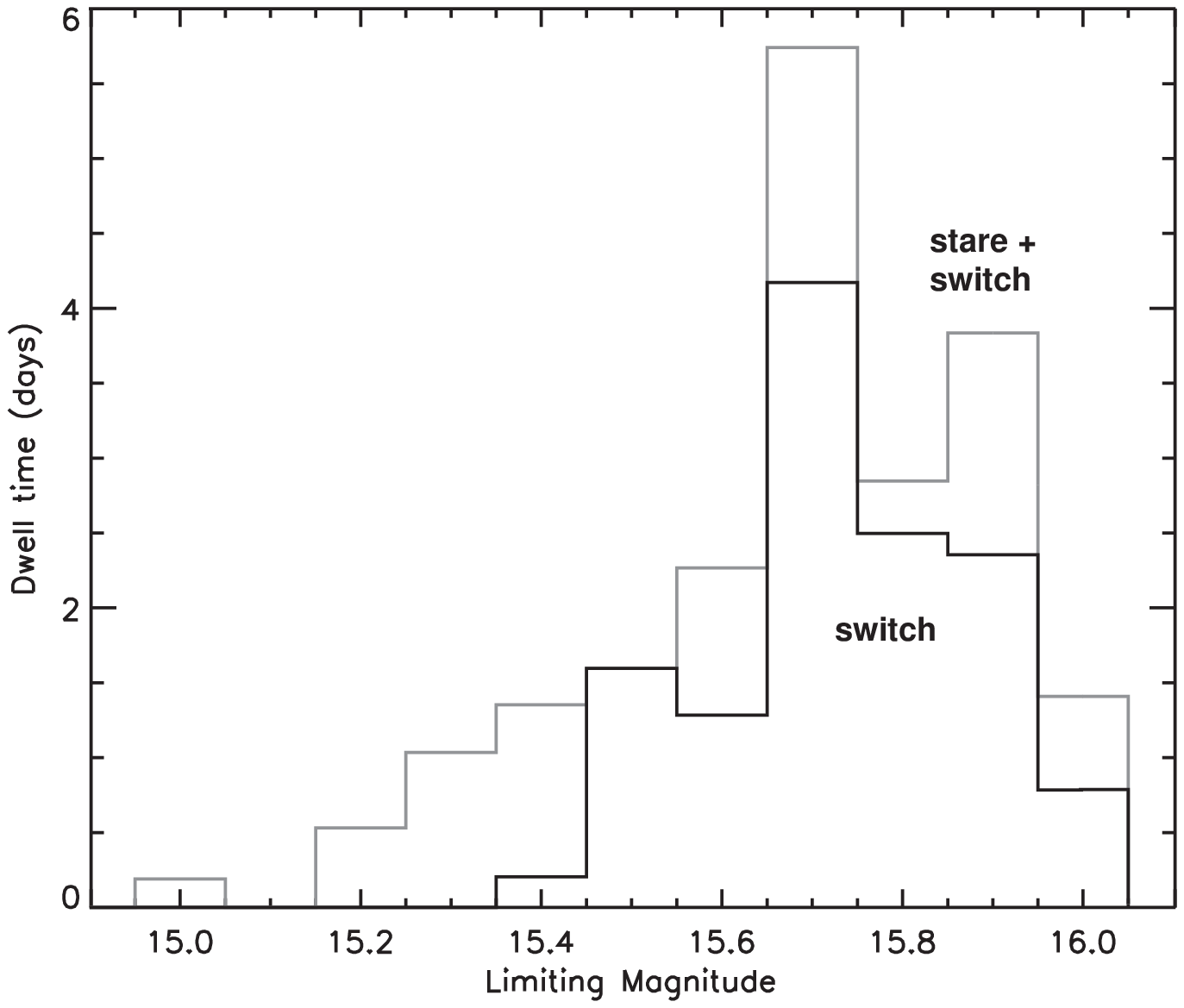}} \par}
\vspace{0.3cm}
\caption[data_sens_summary.ps]{
Distribution of integrated observing times vs. limiting magnitude for the 64
square degree field of a single camera. The limiting magnitudes range from 15.0
to 16.1 for the \textit{stare} data, and 15.4 to 16.1 for the \textit{switch}
data. The total observing time for this size field is 20.8 days.}
\label{fig:data_sens}
\end{figure}

\clearpage

\begin{figure}
\vspace{0.3cm}
{\par\centering \resizebox*{18cm}{!}{\includegraphics{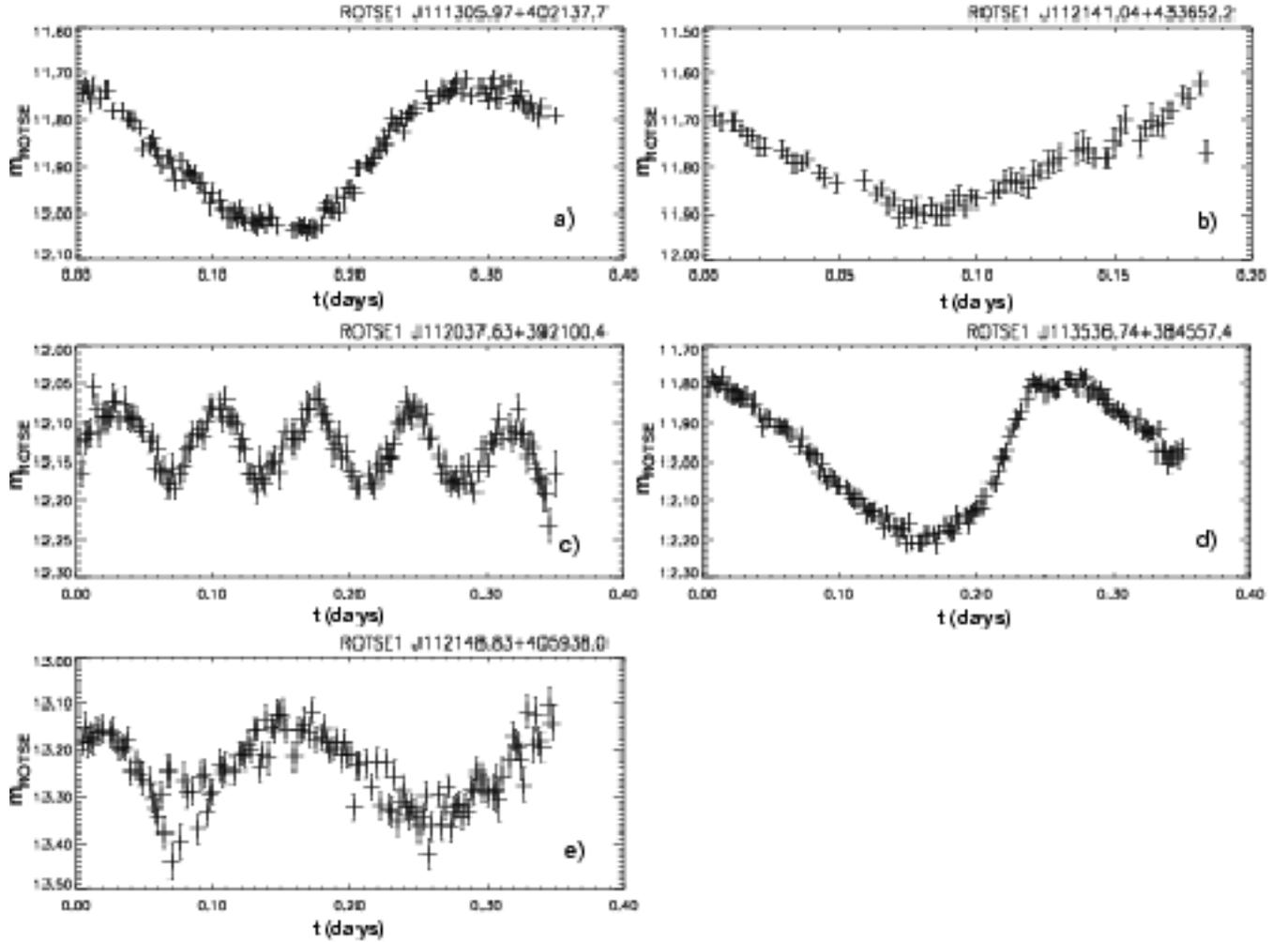}} \par}
\vspace{0.3cm}
\caption[pulsating.ps]{
Five single-night lightcurves for candidate pulsating variables from the April 2000
\textit{stare} data.  None of these are previously identified as variable.  Observations
indicated are good observations after relative photometry.  Errors are stat. $+$ sys.}
\label{fig:pulsing}
\end{figure}

\clearpage

\begin{figure}
\vspace{0.3cm}
{\par\centering \resizebox*{18cm}{!}{\includegraphics{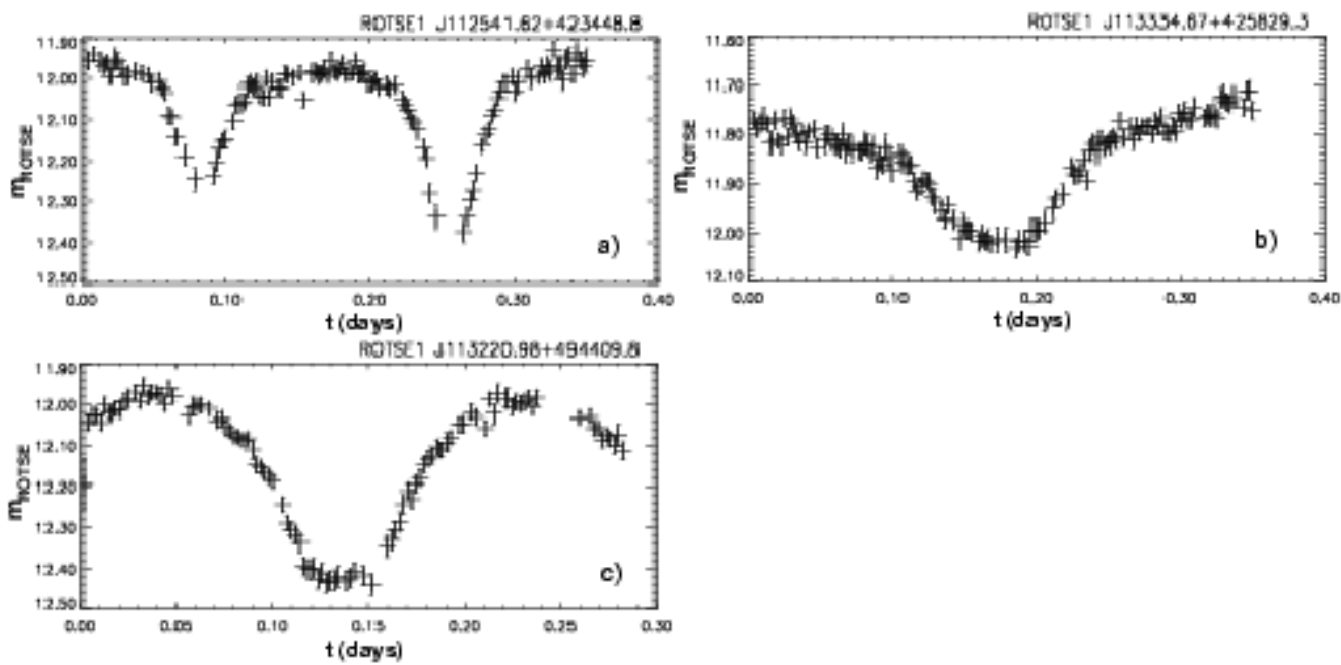}} \par}
\vspace{0.3cm}
\caption[eclipsing.ps]{
Three lightcurves for candidate eclipsing binary variables from the April 2000
\textit{stare} data.  Only ROTSE1 J112541.62+423448.8 was previously known to be 
variable.  Observations indicated are good observations after relative photometry.
Errors are stat. $+$ sys.}
\label{fig:eclipsing}
\end{figure}

\clearpage
\begin{figure}
\vspace{0.3cm}
{\par\centering \resizebox*{10cm}{!}{\includegraphics{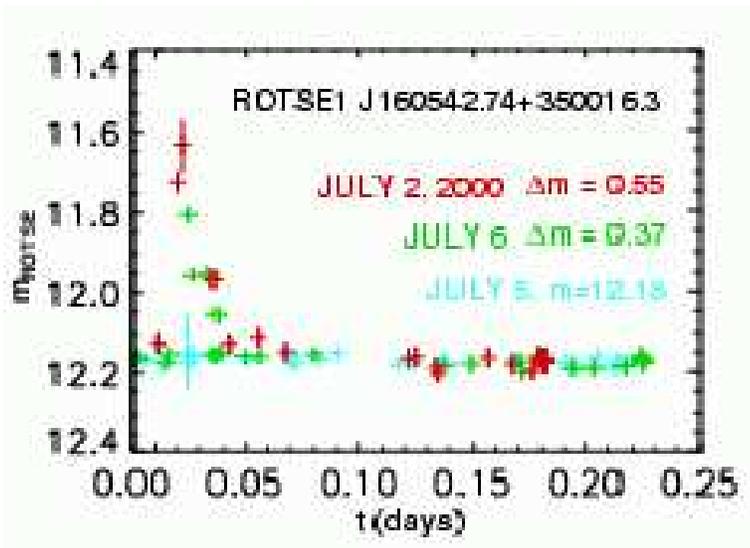}} \par}
\vspace{0.3cm}
\caption[flarestar.ps]{
A candidate flare star from the July 2000 test \textit{switch} data.  Observations
indicated are good observations after relative photometry.  Errors are stat. $+$ sys.
Two outbursts were observed 4 days apart with amplitudes of approximately 0.5 mag 
and durations of about 30 min.  The rest of the lightcurve, including the intermediate 
day July 5, show the source at a constant 12.18 m.  The lightcurves are arbitrarily 
aligned to peak at the same time.}
\label{fig:flaring}
\end{figure}

\clearpage

\begin{figure}
\vspace{0.3cm}
{\par\centering \resizebox*{10cm}{!}{\includegraphics{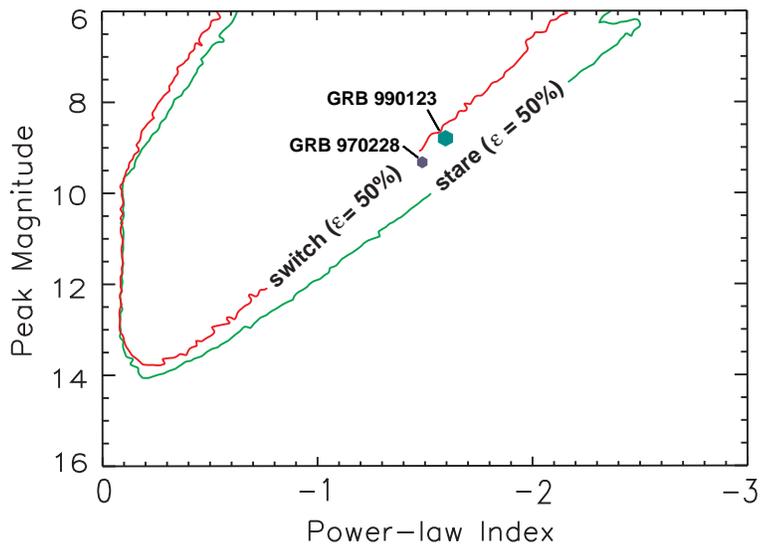}} \par}
\vspace{0.3cm}
\caption[peak_vs_index_compare.ps]{
Sensitive region for observing protocols. The red outline delineates
50\% efficiency for \textit{switch} data, while the green outline indicates
the \textit{stare} data 50\% efficiency. The efficiency turn-on for a 
particular limiting magnitude is quite steep.  Both are calculated for 
co-add image sensitivies of 15.7. The increased efficiency in \textit{stare} 
data for steeply decaying lightcurves is indicated.  The locations of GRB 970228, 
extrapolated from afterglow data \cite{galama00}, and GRB 990123 \cite{akerlof99} 
are shown for comparison.}
\label{fig:mode_sens}
\end{figure}

\clearpage

\begin{figure}
\vspace{0.3cm}
{\par\centering \resizebox*{10cm}{!}{\includegraphics{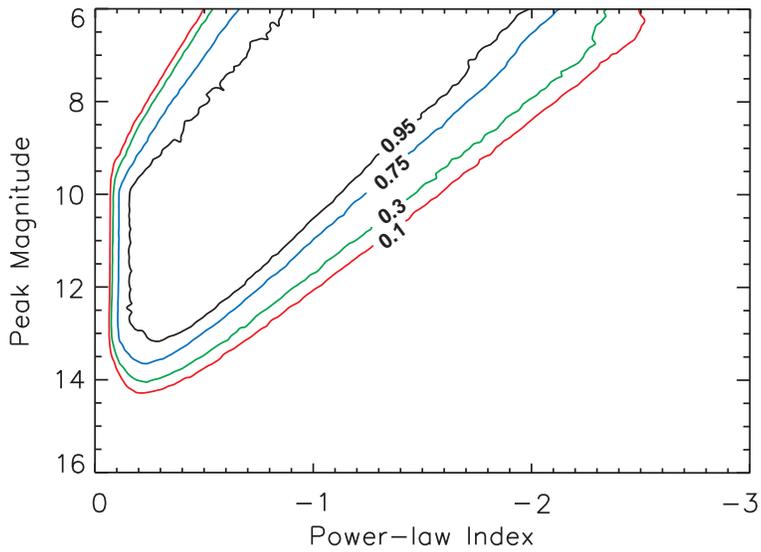}} \par}
\vspace{0.3cm}
\caption[peak_vs_index_all_final.ps]{
Cumulative sensitivity to simulated sources of given \protect\( m_{p}\protect \)
and \protect\( \alpha _{t}\protect \). Contours are generated by folding the
observing time as given in Figure 1 with the protocol efficiency contours such
as given in Figure 5. Contours giving the 95\%, 75\%, 30\% and 10\% efficiency
levels are specified. The gradient in efficiency is primarily due to the varying
protocol and limiting magnitudes of the ROTSE-I observations.
The efficiencies are given relative to a rate of 
\protect\( 1.1\times 10^{-8}~{\rm s}^{-1}~{\rm deg}^{-2}\protect \).}
\label{fig:search_result}
\end{figure}

\clearpage

\begin{table}
  \begin{center}
  \begin{tabular}{|c|c|c|c||c|c||c|} 
	year and month  & dates         & \(\alpha\) & \(\delta\)       & \(\Omega (deg^{2}) \) & dwell-time \( \times \) area 		    & pointing mode\\
	\hline
	Dec, 1999 	& 16 	        & 4h   	     & +15\(^\circ  \)   & 768 			& \( 4.355\times 10^{6}{\rm deg}^{2}{\rm s} \)  & stare\\
		  	& 17 	        & 5h24m      & +11.4\(^\circ  \) & 512 			& 				    & stare\\
	Apr, 2000 	& 9,10,13,14    & 11h42m     & +44\(^\circ  \)   & 768 			& \( 3.9375\times 10^{7}{\rm deg}^{2}{\rm s} \) & stare\\
		  	& 15-17         & 	     & 			& (1024)         	& 				    & \\
	Aug, 2000 	& 24,25,31      & 0h 	     & +30\(^\circ  \)   & 1024 			& \( 4.2799\times 10^{7}{\rm deg}^{2}{\rm s} \) & switch\\
	Sep, 2000 	& 1-3,5,6,28,29 & 	     & 			& 			& 				    & \\
	Oct, 2000 	& 1-4,6 	& 	     & 			& 			& 				    & \\
	Aug, 2000 	& 24,25,31 	& 23h 	     & +15\(^\circ  \) 	& 1024 			& \( 2.8488\times 10^{7}{\rm deg}^{2}{\rm s} \) & switch\\
	Sep, 2000 	& 1-3,5,6,29 	& 	     & 			& 			& 				    & \\
	Oct, 2000 	& 1-4 		& 	     & 			& 			& 				    & \\
  \end{tabular}
  \caption{Itemization of observation properties. The coordinates of fields 
	and total area covered are given, as well as the dates and observing 
	protocol for each.
	\label{tab:datasum}}
  \end{center}
\end{table}

\end{document}